\documentclass[12pt,a4paper]{book}
\usepackage{makeidx}
\usepackage{sussp}
\usepackage{amsfonts}
\usepackage{amssymb}
\usepackage{graphicx}

\makeindex
\def\fun#1#2{\lower3.6pt\vbox{\baselineskip0pt\lineskip.9pt
  \ialign{$\mathsurround=0pt#1\hfil##\hfil$\crcr#2\crcr\sim\crcr}}}
\def\lap{\mathrel{\mathpalette\fun <}}
\def\gap{\mathrel{\mathpalette\fun >}}
\def\mh{M_{\bullet}}

\begin{document}

\headings{Non-Integrable Galactic Dynamics}
{Non-Integrable Galactic Dynamics}
{David Merritt}
{Rutgers University, New Brunswick, NJ USA}

\vspace{15mm}

\section{INTRODUCTION}

Galaxies\index{galaxies} have traditionally been viewed as integrable
or nearly integrable systems, in which the majority of stellar
orbits are regular, respecting as many integrals of motion
as there are degrees of freedom.
Three arguments have commonly been cited in support of this view.
First, many reasonable potentials 
contain only modest numbers of stochastic orbits
\footnote{The terms ``stochastic'' and ``chaotic'' will be used
interchangeably here.}.
This is always true for the potentials of rotationally symmetric 
models,
and there is even a class of non-axisymmetric potentials for which the 
motion is globally integrable, including the famous ``perfect ellipsoid'' (Kuzmin 1973; de Zeeuw \& Lynden-Bell 1985).
Second, stochastic orbits often behave in ways that are very similar 
to regular orbits over astronomically interesting time scales.
Therefore (it is argued) one need not make a sharp distinction 
between regular and stochastic orbits when constructing an 
equilibrium model.
Third, following the successful construction by Schwarzschild 
(1979, 1982) of self-consistent triaxial equilibria, it 
has generally been assumed that the regular orbits -- which are 
confined to narrow regions of phase space and thus have 
definite shapes -- are the fundamental building blocks of 
real galaxies.

Schwarzschild's discovery that many orbits in non-axisymmetric 
potentials are effectively regular came as a surprise, since
triaxial\index{triaxiality} potentials admit only one classical 
integral of the motion, the energy.
In fact a modest fraction of the orbits\index{orbits} 
in Schwarzschild's models were subsequently shown to be stochastic 
(Merritt 1980; Goodman \& Schwarzschild 1981), though only weakly.
But it was clear early on that certain modifications of Schwarzschild's 
potential could lead to a much larger fraction of chaotic orbits.
For instance, Gerhard \& Binney (1985) showed that the addition
of a central density cusp or ``black hole'' (i.e. point mass) to
an otherwise integrable triaxial model would render most of the
center-filling, box orbits unstable, due to deflections that
occur when a trajectory comes close to the center.
This insight was followed by the discovery 
(Crane et al. 1993; Ferrarese et al. 1994)
that stellar spheroids generically contain power-law cusps in the 
luminosity density rather than constant-density cores.
Evidence for central supermassive black holes\index{black holes}
also gradually accumulated (Kormendy \& Richstone 1995).
It is now believed -- not only that black holes are universal components
 of galactic nuclei\index{galactic nuclei} -- 
but that their masses are predictable with
high precision given the global properties of their host spheroids
(Ferrarese \& Merritt 2000; Merritt \& Ferrarese 2001).
Thus it is no longer possible to discuss galaxy dynamics 
in terms of idealized models like Schwarzschild's 
with finite central densities.

Non-integrability has two important consequences.
First, some orbits in non-integrable potentials respect fewer isolating 
integrals than there are degrees of freedom.
Such orbits are chaotic and behave in ways that are very different
from regular orbits: they are exponentially unstable to small
perturbations, and occupy a phase-space region of larger dimensionality
than the invariant tori of regular orbits.
The time-averaged shape of a chaotic orbit is similar to that of an 
equipotential surface and hence such orbits are much less ``useful'' 
than regular orbits for reinforcing the shape of the galaxy's figure.
Second, while regular orbits generally still exist in potentials that 
are not globally integrable, they are strongly influenced by 
resonances\index{resonances} 
between the frequencies of motion in different directions.
These resonances are present even in globally integrable potentials but
have no effect on the structure of phase space;
in non-integrable potentials, however, the resonances divide up
phase space into alternating regions of regular and 
chaotic motion, 
with the lowest-order resonances ``capturing'' the largest parts of
phase space.
Most regular orbits in non-integrable potentials
can be associated with a definite
resonance and have a shape that reflects the order of the resonance.

This article reviews the following topics:
(1) Torus\index{tori} construction, a set of techniques for characterizing 
regular motion in non-integrable potentials and for detecting
departures from integrability\index{integrability};
(2) resonances and their effect on the structure of orbits;
(3) the orbital content of triaxial potentials with central point masses;
(4) mixing, the process by which the phase-space density of stellar
systems approaches a steady state; and
(5) the relation between chaos\index{chaos} 
in the gravitational $N$-body problem\index{gravitational N-body problem}
and chaos in smooth potentials.

\section{TORUS CONSTRUCTION}
	
In systems with a single degree of freedom, constancy of the energy allows
the momentum variable $p$ to be written in terms of the coordinate variable
$q$ as $H(p,q) = E$, and the dependence of both variables on time follows
immediately from Hamilton's equations.  In general systems with $N \geq 2$
degrees of freedom (DOF), such a solution is generally not possible unless
the Hamilton-Jacobi equation is separable, in which case the separation
constants are isolating integrals of the motion. An isolating integral is a
conserved quantity that in some transformed coordinate system makes $\partial
H/\partial p_i = f(q_i)$, thus allowing the motion in $q_i$ to be reduced to
quadratures. Each isolating integral restricts the dimensionality of the
phase space region accessible to an orbit by one; if there are $N$ such
integrals, the orbit moves in a phase space of dimension $2N - N = N$, and
the motion is regular. The $N$-dimensional phase space region to which a
regular orbit is confined is topologically a torus (Figure 1). 
Orbits in time-independent potentials may be either regular or chaotic;
chaotic orbits respect a smaller number of integrals than $N$ -- 
typically only the energy integral $E$.  
Although chaotic orbits are not confined to tori, 
numerical integrations suggest that many chaotic trajectories are effectively
regular, remaining confined for long periods of time to regions of phase
space much more restricted than the full energy hypersurface.
 
\begin{figure}
[ht]
\centering
\includegraphics[width=9cm, trim=20 0 0 0]{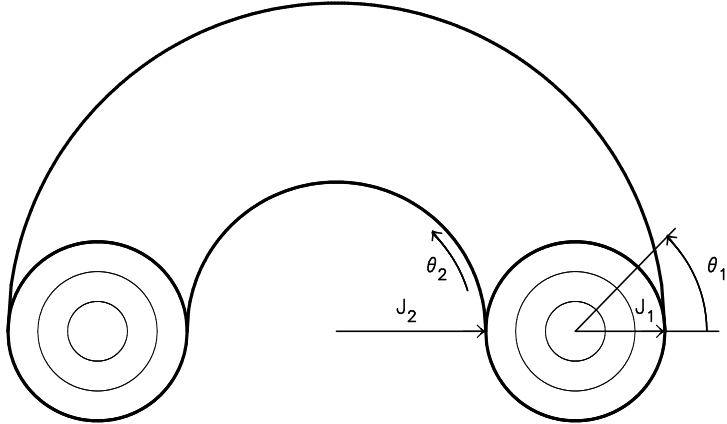}
\caption{
Invariant torus defining the motion of a regular orbit in a
two-dimensional potential.  The torus is determined by the values of the
actions $J_1$ and $J_2$; the position of the trajectory on the torus is
defined by the angles $\theta_1$ and $\theta_2$, which increase linearly with
time, $\theta_i = \omega_it + \theta_i^0$.  
}
\end{figure}
  
The most compact representation of a regular orbit is in terms of the
coordinates on the torus (Figure 1) -- the action-angle variables 
$({\bf J}, \theta)$. The process of determining the map $({\bf x,v}) 
\rightarrow
({\bf J},\theta)$ is referred to as {\it torus construction}. There are a number
of contexts in which it is useful to know the $({\bf J},\theta)$. One example is
the response of orbits to slow changes in the potential, which leave the
actions (${\bf J}$) unchanged.  Another is the behavior of weakly chaotic orbits,
which may be approximated as regular orbits that slowly diffuse from one
torus to another.  A third example is galaxy modeling, where regular orbits
are most efficiently represented and stored via the coordinates that define
their tori.
 
Two general approaches to torus construction have been developed.  
Trajectory-following algorithms are
based on the quasi-periodicity of regular motion: Fourier decomposition of
the trajectory yields the fundamental frequencies on the torus as well as the
spectral amplitudes, which allow immediate construction of the map $\theta
\rightarrow {\bf x}$ in the form of a Fourier series. 
Iterative approaches begin from some initial guess for
${\bf x}(\theta)$, which is then refined via Hamilton's equations with the
requirement that the $\theta_i$ increase linearly with time.  The two
approaches are often complementary, as discussed below.
 
\subsection{Regular Motion}
In certain special potentials, every orbit is regular; examples are the
Kepler and St\"ackel potentials. 
Motion in such globally-integrable potentials can be expressed
most simply by finding a canonical transformation to coordinates $({\bf p,q})$
for which the Hamiltonian is independent of ${\bf q}$, $H=H({\bf p})$; among all such
coordinates, one particularly simple choice is the action-angle variables
$(J_i,\theta_i)$, in terms of which the equations of motion are
\begin{eqnarray}
J_i & = & \hbox{constant,}\nonumber \\
\theta_i & = & \omega_it + \theta_i^0,\ \ \ \ \omega_i= {\partial
H\over\partial J_i}, \ \ \ i = 1,...,N
\end{eqnarray}
(Landau \& Lifshitz 1976; Goldstein 1980).  The trajectory 
${\bf x}({\bf J},\theta)$
is periodic in each of the angle variables $\theta_i$, which may be
restricted to the range $0<\theta_i\le 2\pi$. The $J_i$ define the
cross-sectional areas of the torus while the $\theta_i$ define the position on
the torus (Figure 1).  These tori are sometimes called ``invariant'' since a
phase point that lies on a torus at any time will remain on it forever.
 
Most potentials are not globally integrable, 
but regular orbits may still exist;
indeed these are the orbits for which torus construction machinery is
designed.  
One expects that for a regular orbit in a non-integrable
potential, a canonical transformation $({\bf x,v}) \rightarrow 
({\bf J},\theta)$
can be found such that
\begin{equation}
\dot J_i = 0,\ \ \ \ \dot \theta_i = \omega_i, \ \ \ i = 1,...,N.
\end{equation} 
However there is no guarantee that the full Hamiltonian will
be expressible as a continuous function of the $J_i$ as in
globally integrable potentials.
In general, the map $({\bf x}, {\bf v}) \rightarrow ({\bf J}, \theta)$ w
ill be different for each orbit and
will not exist for those trajectories that do not respect $N$ isolating
integrals.

The uniform translation of a regular orbit on its torus implies that the
motion in any canonical coordinates $({\bf x}, {\bf v})$ is
quasi-periodic\index{quasi-periodicity}:
\begin{eqnarray}
{\bf x}(t) & = &\sum_k {\bf X}_k({\bf J}) \exp\left[ i\left(l_k\omega_1 + m_k\omega_2 +
n_k\omega_3\right)t\right], \nonumber \\
{\bf v}(t) & = &\sum_k{\bf V}_k({\bf J}) \exp\left[ i\left(l_k\omega_1 + m_k\omega_2 +
n_k\omega_3\right)t\right],
\label{quasip}
\end{eqnarray}
with $(l_k,m_k,n_k)$ integers.  The Fourier transform of ${\bf x}(t)$ or 
${\bf v}(t)$
will therefore consist of a set of spikes at discrete frequencies
$\omega_k=l_k\omega_1 + m_k\omega_2 + n_k\omega_3$ that are linear
combinations of the $N$ fundamental frequencies $\omega_i$, with spectral
amplitudes ${\bf X}_k({\bf J})$ and ${\bf V}_k({\bf J})$.
 
\subsection{Trajectory-Following Approaches}
The most straightforward, and probably the most robust, approach to torus
construction is via Fourier analysis of the numerically-integrated
trajectories (Percival 1974; Boozer 1982; Binney \& Spergel 1982, 1984;
Kuo-Petravic et al. 1983; Eaker et al. 1984; Martens \& Ezra 1985).  The
Fourier decomposition of a quasiperiodic orbit (equation \ref{quasip}) 
yields a discrete
frequency spectrum.  The precise form of this spectrum depends on the
coordinates in which the orbit is integrated, but certain of its properties
are invariant, including the $N$ fundamental frequencies $\omega_i$ from
which every line is made up, $\omega_k=l_k\omega_1 + m_k\omega_2 +
n_k\omega_3$.  Typically the strongest line in a spectrum lies at one of the
fundamental frequencies; once the $\omega_i$ have been identified, the
integer vectors $(l_k,m_k,n_k)$ corresponding to every line $\omega_k$ are
uniquely defined, to within computational uncertainties. Approximations to
the actions may then be computed using Percival's (1974) formulae; e.g. the
action associated with $\theta_1$ in a 3 DOF system is
\begin{equation} J_1 = \sum_k
l_k\left(l_k\omega_1+m_k\omega_2+n_k\omega_3\right) |{\bf X}_k|^2 \label{perc1}.
\end{equation} 
Finally, the maps
$(\theta \rightarrow {\bf x})$ are obtained by making the substitution
$\omega_it\rightarrow\theta_i$ in the  spectrum, e.g.
\begin{eqnarray} x(t) & = &\sum_kX_k(J) \exp\left[ i\left(l_k\omega_1 +
m_k\omega_2 + n_k\omega_3\right)t\right] \nonumber \\ & = &\sum_kX_k(J)
\exp\left[ i\left(l_k\theta_1 + m_k\theta_2 + n_k\theta_3\right)\right]
\nonumber \\ & = & x(\theta_1, \theta_2, \theta_3).
\end{eqnarray} 
Trajectory-following algorithms are easily automated; for
instance, integer programming may be used to recover the vectors
$(l_k,m_k,n_k)$ (Valluri \& Merritt 1998).
 
Binney \& Spergel (1982) pioneered the use of trajectory-following algorithms
for galactic potentials. They integrated orbits for a time $T$ and computed
discrete Fourier transforms, yielding spectra in which each frequency spike
was represented by a peak with finite width $\sim\pi/T$ centered on
$\omega_k$. They then fitted these peaks to the expected functional form
$X_k\sin[(\omega-\omega_k)T]/(\omega-\omega_k)$ using a least-squares
algorithm. They were able to recover the fundamental frequencies in a 2 DOF
potential with an accuracy of $\sim 0.1\%$ after $\sim 25$ orbital periods.
Binney \& Spergel (1984) used equation (4) to construct the ``action map''
for orbits in a principal plane of the triaxial logarithmic potential.
Carpintero \& Aguilar (1998) have 
applied similar algorithms to motion in 2- and 3 DOF potentials.
 
The accuracy of Fourier transform methods can be greatly improved by
multiplying the time series with a windowing function before transforming.
The result is a reduction in the amplitude of the side lobes of each
frequency peak at the expense of a broadening of the peaks; the amplitude
measurements are then effectively decoupled from any errors in the
determination of the frequencies.  Laskar (1988, 1990) developed this idea
into a set of tools, the ``numerical analysis of fundamental frequencies''
(NAFF), which he applied to the analysis of weakly chaotic motion in the
solar system.  Laskar's algorithm recovers the fundamental frequencies with
an error that falls off as $T^{-4}$ (Laskar 1996), compared with $\sim
T^{-1}$ in algorithms like Binney \& Spergel's (1982).  Even for modest
integration times of $\sim 10^2$ orbital periods, the NAFF algorithm is able
to recover fundamental frequencies with accuracies of $\sim10^{-8}$ or better
in many potentials. The result is a very precise representation of the torus
(Figure 2).
 
\begin{figure}
[ht]
\centering
\includegraphics[width=12cm, trim=20 0 0 0]{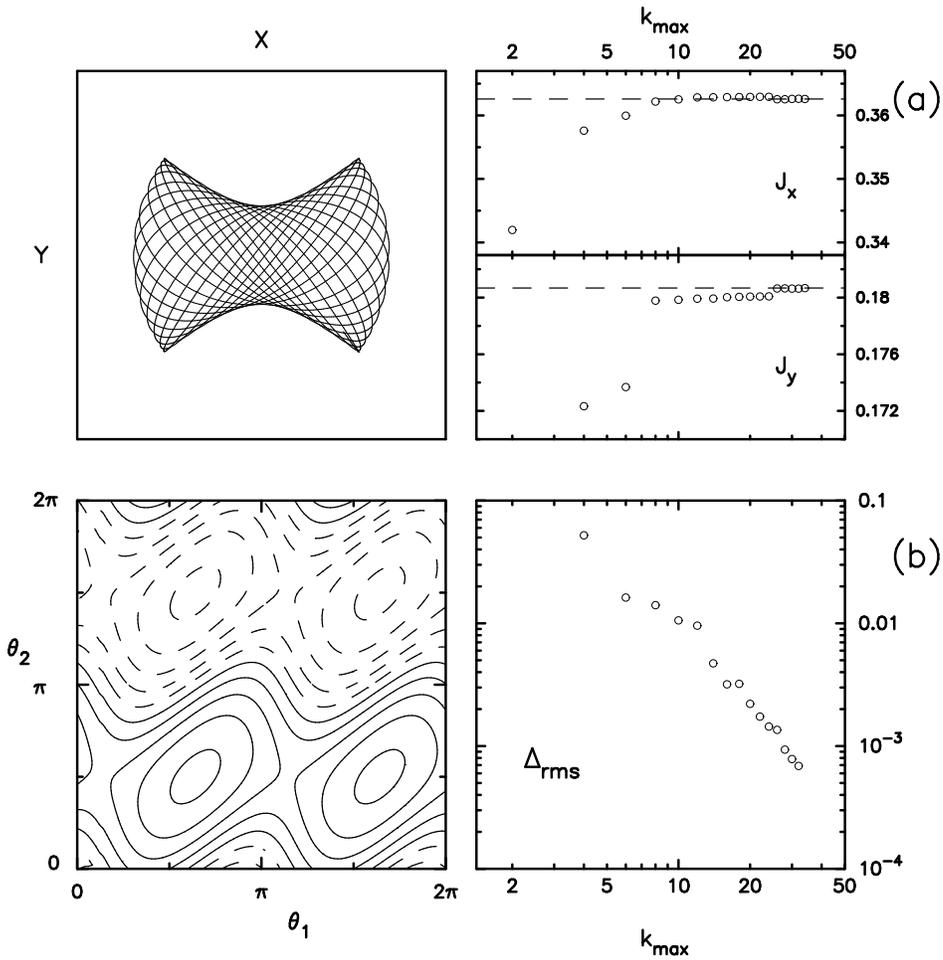}
\caption{
Construction of a 2 DOF, box-orbit torus in a St\"ackel potential
using the NAFF trajectory-following algorithm.  (a) The orbit and its
actions, computed using equation (4) with $k_{max}$ terms. Dashed lines show
the exact $J_i$.  (b) The map $y(\theta_1,\theta_2)$; 
dashed contours correspond to negative values of $y$. $\Delta(k_{max})$ 
is the RMS error in the reconstructed map, 
calculated using an equation similar to (5). 
}
\end{figure}

Since Fourier techniques focus on the frequency domain, they are particularly 
well suited to identifying regions of phase space associated with resonances.
Resonant tori are places where perturbation expansions of integrable systems
break down, due to the ``problem of small denominators''. In perturbed 
(non-integrable) potentials, one expects stable resonant tori to generate 
regions of regular motion and unstable resonant tori to give rise to chaotic
regions.  Algorithms like NAFF allow one to construct a ``frequency map'' 
of the phase space: a plot of the ratios of the fundamental frequencies
$(\omega_1/\omega_3,\omega_2/\omega_3)$ for a large a set of orbits selected
from a uniform grid in initial condition space. Resonances appear on the
frequency map as lines, either densely filled lines in the case of stable
resonances, or gaps in the case of unstable resonances; the frequency map
is effectively a representation of the Arnold web (Laskar 1993).
Resonances are discussed in more detail in $\S 3$.

\subsection{Iterative Approaches}
 
Iterative approaches to torus construction consist of finding successively
better approximations to the map $\theta \rightarrow {\bf x}$ given some 
initial guess ${\bf x}(\theta)$; canonical perturbation theory is a special 
case, and in fact iterative schemes often reduce to perturbative methods in 
appropriate limits. 
Iterative algorithms were first developed in the context of
semi-classical quantization for computing energy levels of bound molecular
systems, and they are still best suited to assigning energies to actions,
$H({\bf J})$. Most of the other quantities of interest to galactic dynamicists -- e.g. the fundamental frequencies $\omega_i$ -- are not easily recovered 
using these algorithms. Iterative schemes also tend to be numerically
unstable unless the initial guess is close to the true solution. On the other
hand, iterative algorithms can be more efficient than trajectory-following
algorithms for orbits that are near (but not exactly on) resonances.
 
Ratcliff, Chang \& Schwarzschild (1984) pioneered iterative schemes in
galactic dynamics. They noted that the equations of motion of a 2 DOF regular
orbit,
\begin{equation} \ddot x = -{\partial\Phi\over\partial x},\ \ \ \ \ddot y =
-{\partial\Phi\over\partial y},
\end{equation} can be written in the form
\begin{eqnarray} \left(\omega_1{\partial\over\partial\theta_1} +
	\omega_2{\partial\over\partial\theta_2}\right)^2x & = &
	-{\partial\Phi\over\partial x}, \nonumber \\
\left(\omega_1{\partial\over\partial\theta_1} +
	\omega_2{\partial\over\partial\theta_2}\right)^2y & = &
	-{\partial\Phi\over\partial y}.
\label{rat1}
\end{eqnarray}
If one specifies $\omega_1$ and $\omega_2$ and treats $\partial\Phi/\partial
x$ and $\partial\Phi/\partial y$ as functions of the $\theta_i$, equations
(\ref{rat1}) can be viewed as nonlinear differential equations for
$x(\theta_1,\theta_2)$ and $y(\theta_1,\theta_2)$. Ratcliff et al. expressed
the coordinates as Fourier series in the angle variables,
\begin{equation}
{\bf x}(\theta) = \sum_{n} {\bf X}_{n} e^{i{n}\cdot{\theta}}.
\label{rat2}
\end{equation}
Substituting (\ref{rat2}) into (\ref{rat1}) gives
\begin{equation} \sum_{n}({\bf n}\cdot\omega)^2{\bf X}_{n}e^{i{n \cdot
\theta}}=\nabla\Phi
\label{rat4}
\end{equation}
where the right hand side is again understood to be a function of the angles.
Ratcliff et al. truncated the Fourier series after a finite number of terms
and required equations (\ref{rat4}) to be satisfied on a grid of points around the
torus. They then solved for the ${\bf X}_n$ by iterating from an initial guess.
Convergence was found to be possible if the initial guess was close to the
exact solution. A similar algorithm was developed for recovering tori in the
case that the actions, rather than the frequencies, are specified a priori.
Guerra \& Ratcliff (1990) applied these algorithms to motion in the plane of
rotation of a nonaxisymmetric potential.
 
Another iterative approach to torus construction was developed by Chapman,
Garrett \& Miller (1976) in the context of semiclassical quantum theory. One
begins by dividing the Hamiltonian $H$ into separable and non-separable parts
$H_0$ and $H_1$, then seeks a generating function $S$ that maps the known
tori of $H_0$ into tori of $H$. For a generating function of the $F_2$-type
(Goldstein 1980), one has
\begin{equation}
{\bf J}(\theta,{\bf J}') = {\partial S\over\partial\theta}, \ \ \ \
\theta'(\theta,{\bf J}') = {\partial S\over\partial {\bf J}'}
\label{gold}
\end{equation}
where $({\bf J}, \theta)$ and $({\bf J}', \theta')$ are the action-angle variables
of $H_0$ and $H$ respectively. The generator $S$ is determined, for a
specified ${\bf J}'$, by substituting the first of equations (\ref{gold}) into the
Hamiltonian and requiring the result to be independent of
$\theta$. One then arrives at $H({\bf J}')$. Chapman et al. showed that a
sufficiently general form for $S$ is
\begin{equation}
S(\theta,{\bf J}') = \theta \cdot {\bf J}' - i\sum_{{n}\ne 0}
S_{n}({\bf J}')e^{i{n}\cdot\theta},
\end{equation}
where the first term is the identity transformation, and they evaluated a
number of iterative schemes for finding the $S_{n}$. 
One such scheme was
found to recover the results of first-order perturbation theory after a
single iteration. 
McGill \& Binney (1990) applied the Chapman et al.
algorithm to 2 DOF motion in the axisymmetric logarithmic
potential.
 
The generating function approach is not naturally suited to deriving the
other quantities of interest to galactic dynamicists. For instance, equation
(\ref{gold}) gives $\theta'(\theta)$ as a derivative of $S$, but since $S$ must be
computed separately for every ${\bf J}'$ its derivative is likely to be
ill-conditioned. Binney \& Kumar (1993) and Kaasalainen \& Binney (1994a)
discussed two schemes for finding $\theta'(\theta)$; the first requires the
solution of a formally infinite set of equations, while the latter requires
multiple integrations of the equations of motion for each torus --
effectively a trajectory-following scheme.
 
Warnock (1991) presented a hybrid scheme in which the generating function $S$
was derived by numerically integrating an orbit from appropriate initial
conditions, transforming the coordinates to $({\bf J}, \theta)$ of $H_0$ and
interpolating ${\bf J}$ on a regular grid in $\theta$. The values of the
$S_{n}$ then follow from the first equation of (\ref{gold}) after a discrete
Fourier transform. Kaasalainen \& Binney (1994b) found that Warnock's scheme
could be used to substantially refine the solutions found via their iterative
algorithm. Another hybrid scheme was discussed by Reiman \& Pomphrey (1991).
 
Having computed the energy on a grid of ${\bf J}'$ values, one can interpolate to
obtain the full Hamiltonian $H({\bf J}')$. If the system is not in fact
completely integrable, this $H$ may be rigorously interpreted as smooth
approximation to the true $H$ (Warnock \& Ruth 1991, 1992) and can be taken
as the starting point for secular perturbation theory. Kaasalainen (1994)
developed this idea and showed how to recover accurate surfaces of section in
the neighborhood of low-order resonances in the planar logarithmic potential.
 
Percival (1977) described a variational principle for constructing tori. His
technique has apparently not yet been implemented in the context of galactic
dynamics.
 
\subsection{Chaotic Motion}
 
Torus-construction machinery may be applied to orbits that are approximately,
but not precisely, regular (Laskar 1993).  The frequency spectrum of a weakly
chaotic orbit will typically be close to that of a regular orbit, with most
of the lines well approximated as linear combinations of three ``fundamental
frequencies'' $\omega_i$. 
However these frequencies will change with time as
the orbit migrates from one ``torus'' to another.  
The diffusion rate can be
measured via quantities like $\delta\omega=|\omega_1 - \omega'_1|$, 
the change in a
``fundamental frequency'' over two consecutive integration intervals.
Papaphilippou \& Laskar (1996, 1998), Valluri \& Merritt (1998) and Wachlin
\& Ferraz-Mello (1998) used this technique to study chaos and diffusion in
triaxial galactic potentials. 

Measuring chaos via quantities like $\delta\omega$ has a number of advantages
over the traditional technique based on computation of the Liapunov
exponents\index{Liapunov exponents} (Lichtenberg \& Lieberman 1992).
$\delta\omega$ can be accurately determined after just a few tens of
orbital periods, whereas determination of the Liapunov exponents may 
require much longer integrations.
The Liapunov exponents measure only the rate of growth of infinitesimal
perturbations around the trajectory, 
while $\delta\omega$ measures the finite ``movement'' of the trajectory 
in action-angle space, a more physically interesting measure of chaos.
It is possible for orbits to be extremely unstable in the sense of having
large Liapunov exponents, but to behave nearly regularly in the sense
of having small $\delta\omega$; an example is presented in
$\S 6$.

\section{RESONANCES}

The character of a regular orbit depends critically on whether the
fundamental frequencies $\omega_i$ are independent, or whether they satisfy
one or more nontrivial linear relations of the form
\begin{equation} 
\sum_{i=1}^N m_i\omega_i=0
\label{res1}
\end{equation} 
with $N$ the number of degrees of freedom and $m_i$ integers,
not all of which are zero.
Generally there exists no relation like equation (\ref{res1});
the frequencies are incommensurate, and the trajectory fills its
invariant torus uniformly and densely in a time-averaged sense.
When one or more resonance relations are satisfied, however,
the trajectory is restricted to a phase-space region of lower
dimensionality than $N$.

\begin{figure}
[ht]
\centering
\includegraphics[width=6cm, angle=270, trim=0 0 0 0]{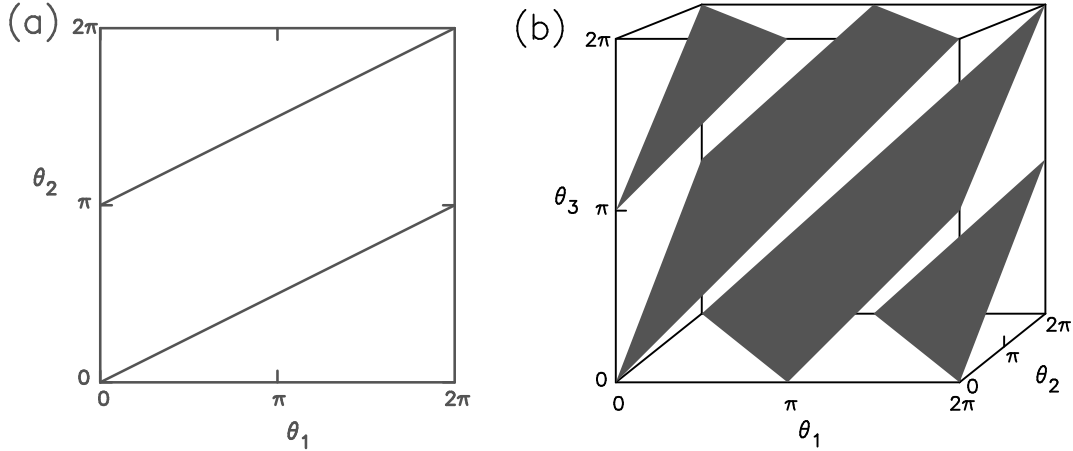}
\caption{
Resonant tori. (a) A two-dimensional torus, shown here as a square
with identified edges. The plotted trajectory satisfies a $2:1$ resonance
between the fundamental frequencies, $\omega_1 - 2\omega_2 = 0$ (e.g. a
``banana''). (b) A three-dimensional torus, shown here as a cube with
identified sides. The shaded region is covered densely by a resonant
trajectory for which $2\omega_1 + \omega_2 - 2\omega_3 = 0$. This trajectory
is not closed, but it is restricted by the resonance condition to a
two-dimensional subset of the torus. The orbit in configuration space is thin.
}
\end{figure}

In the case of a two-dimensional regular orbit,
the angle variables are
\begin{equation}
\theta_1=\omega_1 t + {\theta_1}_0,
\ \ \ \ \theta_2=\omega_2 t + {\theta_2}_0,
\label{reson}
\end{equation}
which define the surface of a torus (Fig. 1).
Because of the quasi-periodicity of the orbit, its torus can be 
mapped onto a square in the $(\theta_1,\theta_2)$-plane, with 
each side ranging from $0$ to $2\pi$ (Figure 3a);
the top and bottom of the square are identified with each other,
as are the left and right sides.
In the general case, the frequencies $\omega_1$ and $\omega_2$
are incommensurate and the trajectory densely covers the entire 
$(\theta_1,\theta_2)$-plane after an infinite time.
However if the ratio $\omega_1/\omega_2=|m_2/m_1|$
is a rational number, i.e. if $m_1$ and $m_2$ are integers,
the orbit closes on itself after 
$|m_2|$ revolutions in $\theta_1$ and $|m_1|$ revolutions in 
$\theta_2$ and fills only a one-dimensional subset of its torus
(e.g. Arnold 1963, p. 164).
Its dimensionality in configuration space is also one -- the orbit
is closed.
Such an orbit has a single fundamental frequency 
$\omega_0 = \omega_1/m_2 = \omega_2/m_1 = 2\pi/T$, with $T$ the 
orbital period; after an elapsed time $T$, the trajectory returns 
to its starting point in phase space.
Examples of resonant orbits in two-dimensional 
galactic potentials are the ``boxlets'' 
(Miralda-Escud\'e \& Schwarzschild 1989).

In the case of a three-dimensional regular orbit, the angle
variables are
\begin{equation}
\theta_1=\omega_1 t + {\theta_1}_0,
\ \ \ \ \theta_2=\omega_2 t + {\theta_2}_0,
\ \ \ \ \theta_3=\omega_3 t + {\theta_3}_0.
\end{equation}
The orbit may now be mapped into a cube whose axes are 
identified with the $\theta_i$ (Figure 3b).
If the $\omega_i$ are incommensurate, this cube will be 
densely filled after a long time.
However if a single condition of the form
\begin{equation}
m_1\omega_1 + m_2\omega_2 + m_3\omega_3 = 0
\label{commen}
\end{equation}
is satisfied with integer $m_i$, 
the motion is restricted for all time to a two-dimensional subset of its torus .
Such an orbit is not closed; instead, as suggested by 
Figure 3b, it is {\it thin}, confined to a 
sheet or membrane in configuration space, which it fills densely
after infinite time.

\begin{figure}
[ht]
\centering
\includegraphics[width=16cm, trim=20 0 0 0]{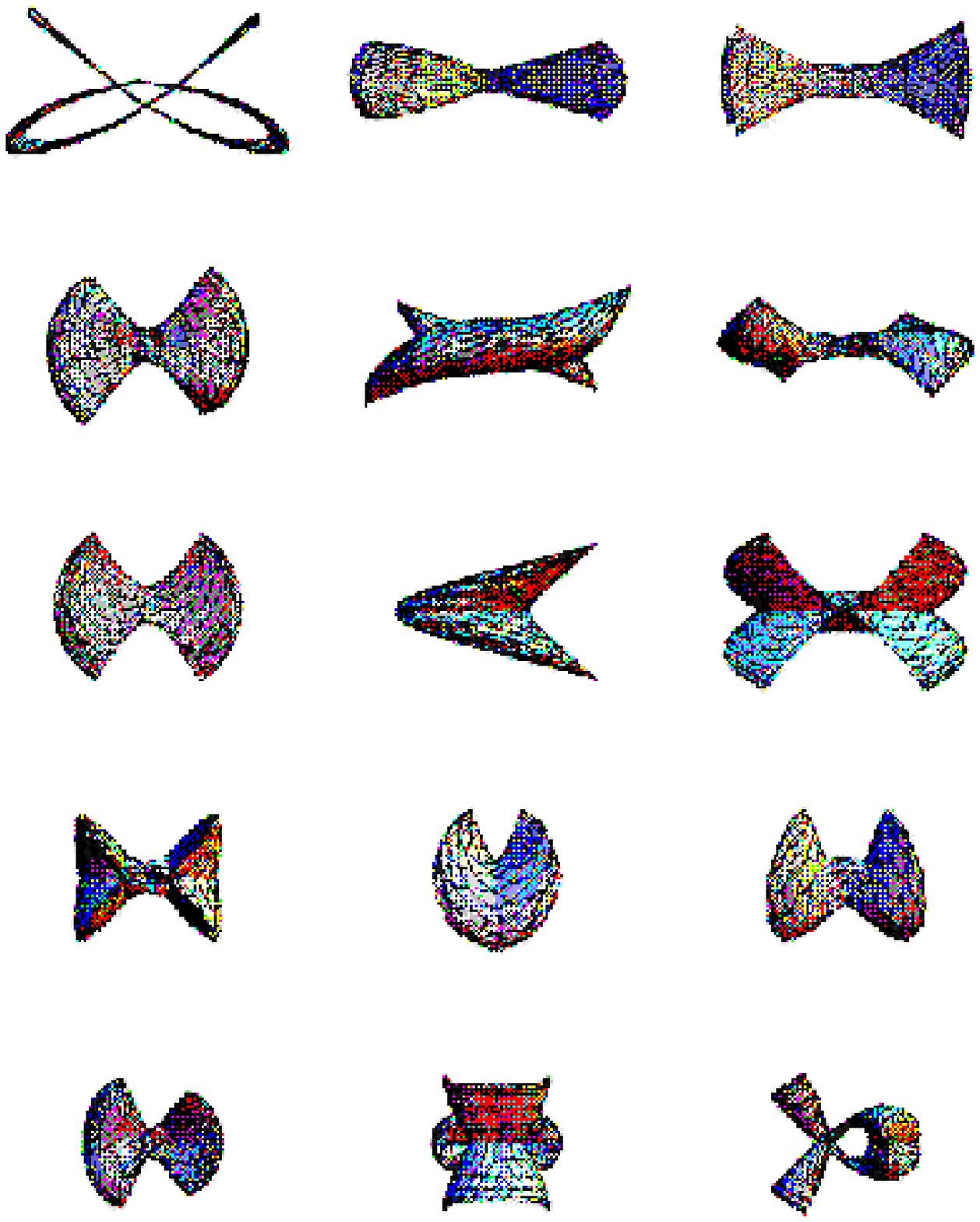}
\caption{ 
Surfaces filled by a set of thin, or resonant, box orbits in
a non-integrable triaxial potential (Merritt \& Valluri 1999),
as seen from vantage points on each of the three principal axes.
The cross sections of these orbits are shown in Figure 5.
}
\end{figure}

\begin{figure}
[ht]
\centering
\includegraphics[width=13cm, trim=20 0 0 0]{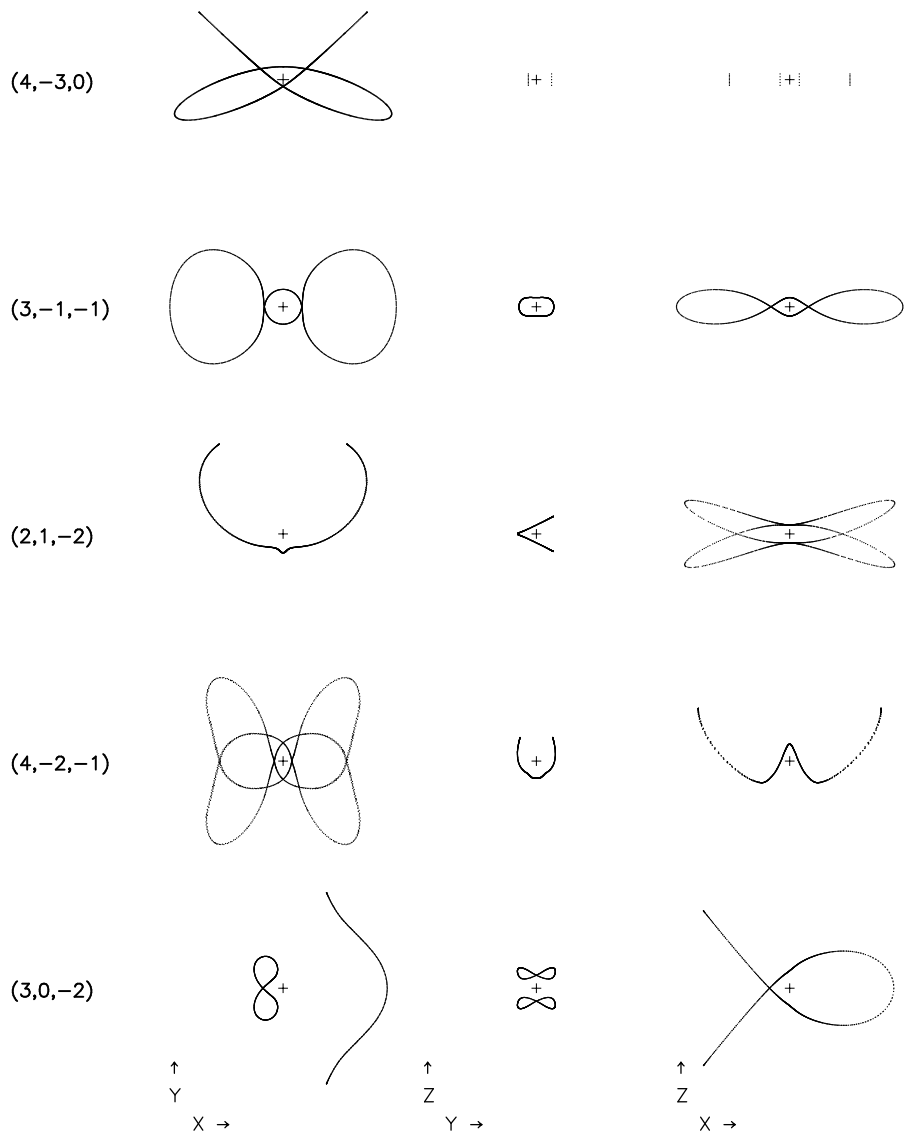}
\caption{ 
Intersections with the principal planes
of the thin box orbits shown in Figure 4.
Because the orbits are thin, their intersections with any plane 
define a curve or set of curves.
The center of the potential is indicated by a cross.}
\end{figure}

Just as in the two-dimensional case,
the condition (\ref{commen}) may be 
used to reduce the number of independent frequencies by one.
Defining the two ``base'' frequencies 
$\omega_0^{(1)}, \omega_0^{(2)}$ as
\begin{equation}
\omega_0^{(1)} = \omega_3/m_1,\ \ \ \ \omega_0^{(2)} = 
\omega_2/m_1,
\end{equation}
we may write
\begin{eqnarray}
\omega_1 & = & -m_3\omega_0^{(1)} - m_2\omega_0^{(2)}, \nonumber \\
\omega_2 & = & m_1\omega_0^{(2)}, \nonumber \\
\omega_3 & = & m_1\omega_0^{(1)}.
\end{eqnarray}
Since the motion is quasi-periodic, i.e.
\begin{equation}
{\bf x}(t)  = \sum_k {\bf X}_k\exp i
\left(l_k\omega_1 + m_k\omega_2 + n_k\omega_3\right)t,
\label{qp1}
\end{equation}
with $(l_k,m_k,n_k)$ integers, it will remain quasi-periodic when 
expressed in terms of the two base frequencies:
\begin{eqnarray}
{\bf x}(t)  & = & \sum_k {\bf X}_k\exp i
\left[\left(-l_k m_3+n_km_1\right)\omega_0^{(1)} + 
\left(-l_k m_2 + m_k m_1\right) \omega_0^{(2)}\right]t \nonumber \\
& = & \sum_k {\bf X}_k\exp i
\left({l_k}'\omega_0^{(1)} + {m_k}'\omega_0^{(2)}\right)t, \nonumber \\ 
& = & \sum_k {\bf X}_k\exp i
\left({l_k}'\theta^{(1)} + {m_k}'\theta^{(2)}\right), \nonumber \\ 
{l_k}' & = & -l_k m_3 + n_k m_1, \ \ \ \ \ {m_k}' = -l_k m_2 + m_k 
m_1, \nonumber \\
\theta^{(1)} & = & \omega_0^{(1)}t, \ \ \ \ \ \ \ \ \ \ \ \ \ \ \ \ \ \ 
\theta^{(2)} = \omega_0^{(2)}t.  
\label{qp2}
\end{eqnarray}
A Fourier transform of the motion will therefore consist of a set 
of spikes whose locations can be expressed as 
linear combinations of just two frequencies.
Equation (\ref{qp2}) is a parametric expression for the Cartesian coordinates
in terms of the angles on the 2-torus, i.e. it is a reconstruction of the 
(reduced) torus.
A number of examples of resonant box orbits reconstructed in this way
are illustrated in Figures 4 and 5.

Certain special orbits may satisfy two independent resonance relations
simultaneously. 
In this case we can write:
\begin{eqnarray}
& m_1\omega_1 + m_2\omega_2 + m_3\omega_3 & = 0, \nonumber \\
& n_1\omega_1 + n_2\omega_2 + n_3\omega_3 & = 0,
\label{degen}
\end{eqnarray}
and each frequency $\omega_i$ may be expressed as a rational fraction 
of any other:
\begin{equation}
{\omega_1\over\omega_3} = {m_2n_3 - m_3n_2\over m_1n_2 - m_2n_1} 
= {l_1\over l_3},\ \ \ \ \ 
{\omega_2\over\omega_3} = {m_3n_1 -m_1n_3\over m_1n_2 - 
m_2n_1} = {l_2\over l_3},
\end{equation}
with $(l_1,l_2,l_3)$ integers.
The motion is therefore periodic with a single base frequency 
$\omega_0=\omega_1/l_1=\omega_2/l_2=\omega_3/l_3$
and the trajectory is closed -- the orbit is a three-dimensional,
closed curve.
In a system with $N$ degrees of freedom, $N-1$ such conditions 
are required for closure; only in the 2DOF case does a single 
resonance condition imply closure.

Following Poincar\'e (1892), it has commonly been assumed that closed
orbits are the fundamental ``building blocks'' of phase space.
However in three-dimensional potentials, 
one expects thin orbits to be more common than closed ones, 
in the sense that orbits satisfying one resonance condition are more likely
than orbits satisfying two.
Hence one expects that most regular orbits will be associated with
families whose parent is a thin orbit.
Numerical integrations of orbits in realistic nonaxisymmetric potentials
suggest that this is in fact the case: the majority of regular orbits
have most of their ``power'' in frequencies that lie close to linear
combinations of two fundamental frequencies (thin orbit) rather than
one frequency (closed orbit) (Merritt \& Valluri 1999; Figure 7).

\begin{figure}
[ht]
\centering
\includegraphics[width=15cm, trim=0 0 0 300]{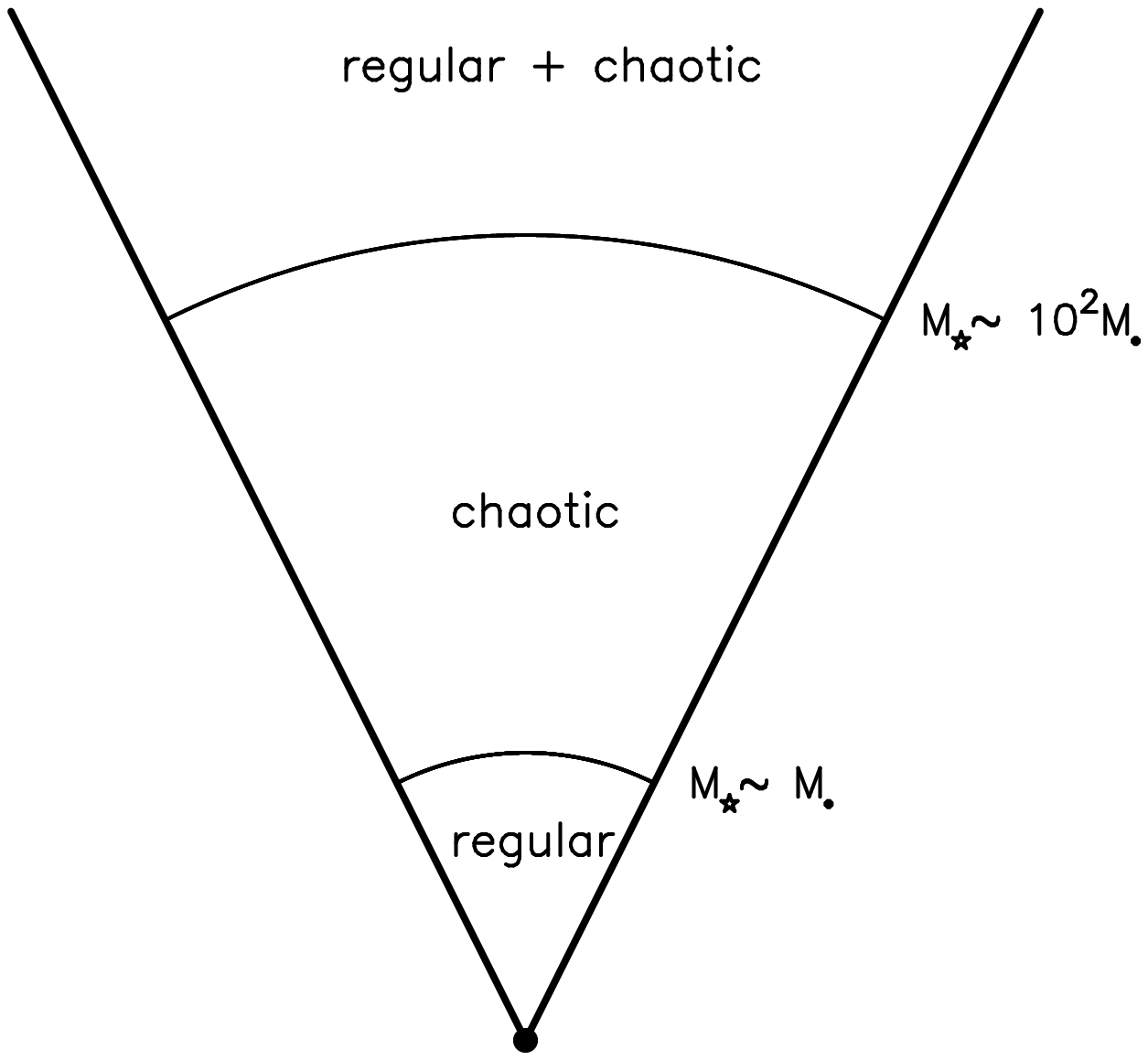}
\caption{ 
Three zones in the phase space of triaxial potentials (see text).
}
\end{figure}

\section{ORBITAL STRUCTURE OF TRIAXIAL POTENTIALS WITH CENTRAL SINGULARITIES}

Non-integrability is likely to be a generic feature of galactic potentials,
for two reasons.
First, galaxies are often observed to be non-axisymmetric, either due
to the presence of imbedded sub-systems like bars, or because the stellar
distribution is globally triaxial.
Observational evidence for global triaxiality in elliptical galaxies
is not particularly strong; 
few ellipticals exhibit significant minor-axis rotation  
(Franx, Illingworth \& de Zeeuw 1991), and detailed modelling of 
a handful of nearby ellipticals suggests that 
their kinematics can often be very well reproduced by assuming
axisymmetry (e.g. van der Marel et al. 1998).
However, at least some elliptical galaxies and bulges exhibit clear 
kinematical signatures of non-axisymmetry (s.g. Schechter \& Gunn 1979;
Franx, Illingworth \& Heckman 1989), and the 
observed distribution of Hubble types is likewise inconsistent
with the assumption that all ellipticals are precisely axisymmetric 
(Tremblay \& Merritt 1995, 1996; Ryden 1996).
Mergers between disk galaxies also produce generically triaxial
systems (Barnes 1996), and departures from axisymmetry (possibly transient) 
are widely argued to be necessary for 
the rapid growth of nuclear black holes during the quasar 
epoch (Shlosman, Begelman \& Frank 1990), 
for the fueling of starburst galaxies (Sanders \& Mirabel 1996),
and for the large radio luminosities of some ellipticals 
(Bicknell et al. 1997).
These arguments suggest that most elliptical galaxies or bulges
may have been triaxial at an earlier epoch, and perhaps
that triaxiality is a recurrent phenomenon induced by 
mergers or other interactions.

The second feature of galactic potentials that is conducive to 
non-integrability
is the apparently universal presence at the centers of stellar
spheroids of high stellar densities and supermassive black
holes.
Low-luminosity ellipticals and bulges have stellar luminosity profiles
that diverge as unbroken power laws at small radii, $\rho\sim r^{-\gamma}$,
with $\gamma\approx 2$.
Brighter galaxies also exhibit power laws in the space density of stars,
but with shallower slopes, $\gamma\lap 1$; seen in projection, these weaker
cusps appear as cores (Kormendy 1985).
The gravitational force in an $r^{-2}$ density cusp diverges as $r^{-1}$,
not steep enough to produce large-angle deflections in the motion of stars
that pass near the center.
However galaxies also contain supermassive black holes,
with masses that correlate astonishingly well with the velocity dispersion
of the stars (Ferrarese \& Merritt 2000);
the ratio of black hole mass to spheroid mass is $\sim 0.0015$ with
small scatter (Merritt \& Ferrarese 2001).
The combination of non-axisymmetry in the potential with a steep
central force gradient is conducive to non-integrability and chaos, 
since many orbits in non-axisymmetric potentials pass near the center 
where they undergo strong gravitational deflections
(Gerhard \& Binney 1985).  

\begin{figure}
[ht]
\centering
\includegraphics[width=11cm, trim=0 0 0 10]{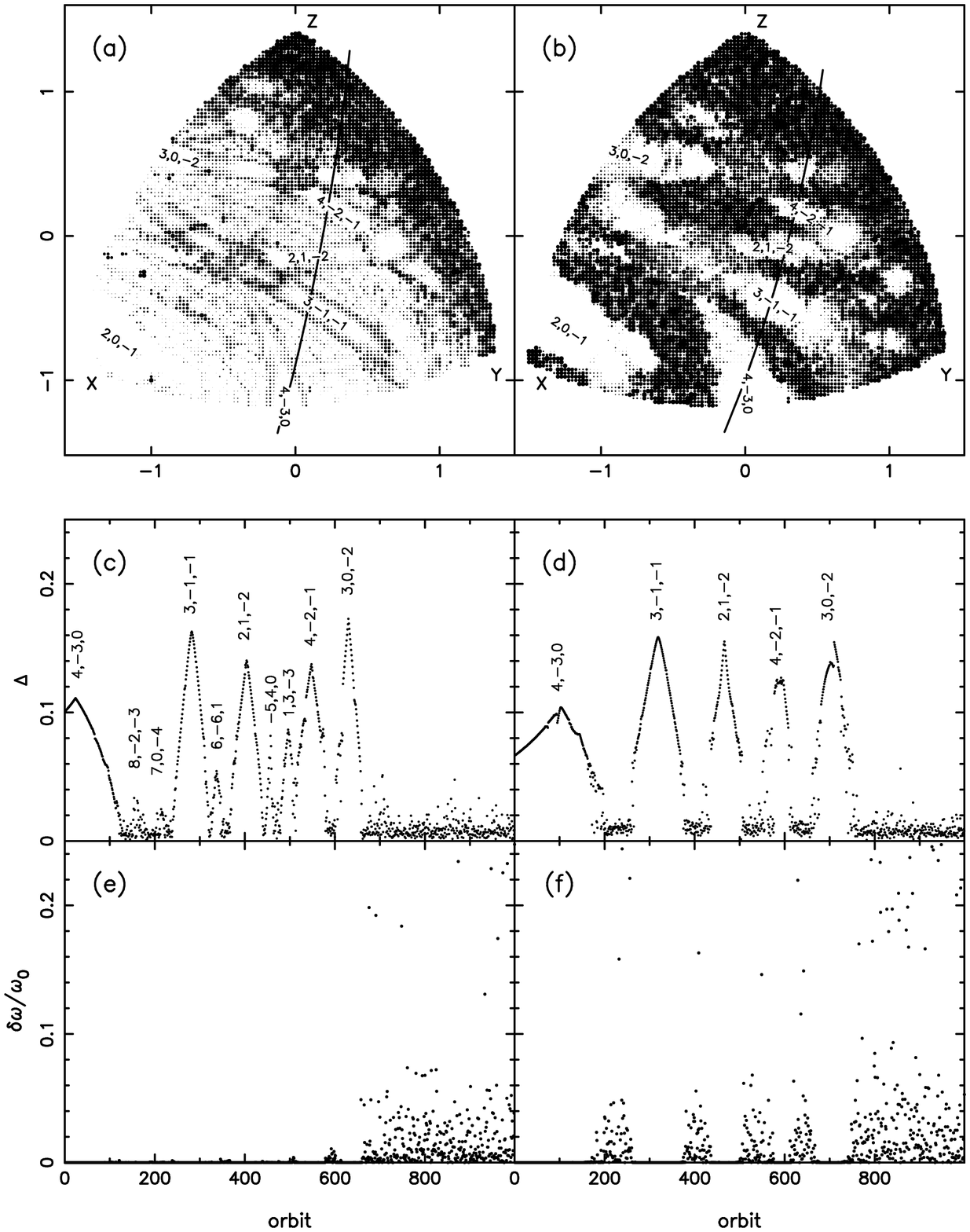}
\caption{ 
Non-integrability in triaxial potentials (Merritt \& Valluri 1999).
The mass model in (a) has a weak ($\gamma=0.5$) density cusp
and no black hole; in (b) the black hole contains $0.3\%$ of the total
mass.
Each panel shows one octant of an equipotential surface, lying close
to the half-mass radius of the model;
the $z$ (short) axis is vertical and the $x$ (long) axis is to the
left.
The grey scale measures the degree of stochasticity of orbits started
with zero velocity on the equipotential surface.
Stable resonance zones -- the white bands in (a) and (b) -- 
are labelled by their defining integers $(m_1,m_2,m_3)$.
Panels (c) and (d) show the pericenter distance $\Delta$ of a set of $10^3$ orbits with starting points along the heavy solid lines in (a) and (b).
Panels (e) and (f) plot a measure of the chaos for these orbits;
$\delta\omega/\omega_0$ is the fractional change in the frequency of the
strongest line in the orbit's frequency spectrum.
}
\end{figure}

In a triaxial potential containing a central point mass, 
the phase space divides naturally into three 
regions depending on energy, i.e. on distance from the center
(Figure 6).
In the innermost region, where the enclosed mass in stars is less
than the mass of the black hole,
the potential is dominated by the central singularity and the motion
is essentially regular.
The gravitaional force from the stars acts as a small perturbation 
causing the nearly-Keplerian orbits around the black hole to slowly 
precess.
The two major orbit families in this region are the tube orbits,
high angular momentum orbits that avoid the center; and the pyramid
orbits, Keplerian ellipses that precess in two orthogonal planes
parallel to the short axis of the figure
(Sridhar \& Touma 1999; Sambhus \& Sridhar 2000; Poon \& Merritt 2001).
Pyramid orbits are similar to the classical box orbits of integrable
triaxial potentials except
that their elongation is counter to that of the
triaxial figure, making them less useful for self-consistently 
reconstructing a galaxy's shape.

At intermediate radii, the black hole acts
as a scattering center rendering almost all of the center-filling or
box orbits stochastic.
(Tube orbits persist at these and higher energies and remain mostly regular.)  
This ``zone of chaos'' extends from a few 
times $r_g$, the radius where the black hole dominates the gravitational
force, 
out to a radius where the enclosed stellar mass is roughly $10^2$ 
times the mass of the black hole. 
The transition to chaos at $r\gap r_g$ is very rapid
and occurs at lower energies in more elongated potentials 
(Poon \& Merritt 2001).

If the black hole mass exceeds $\sim 10^{-2}$ times the mass of
the stellar spheroid, as it may do in a few galaxies 
(Merritt \& Ferrarese 2001), the chaotic zone will include 
essentially the entire potential outside of $\sim r_g$.
However if $M_{\bullet}\approx 10^{-3}M_{\rm gal}$, 
as in the majority of galaxies, 
there exists a third, outermost region where the phase 
space is a complex mixture of chaotic and regular trajectories, including
resonant box orbits like those in Figures 4 and 5 
that remain stable by avoiding the center 
(Carpintero \& Aguilar 1998; Papaphillipou \& Laskar 1998; Valluri \& 
Merritt 1998; Wachlin \& Ferraz-Mello 1998).
Figure 7 illustrates the complexity of box-orbit phase space at large energies
in two triaxial potentials: one with a weak density cusp and the other with
a central point mass.

Non-integrable potentials often exhibit a transition to global 
stochasticity as the magnitude of some perturbation parameter is increased.
The results summarized above suggest that there are two such perturbation
parameters associated with motion in triaxial galaxies containing
central black holes.
In a triaxial galaxy with a given $\mh$, 
the motion of center-filling orbits undergoes a sudden
transition to stochasticity as the energy is increased;
the critical value is the energy at which the gravitational force from
the stars is of order the force from the black hole.
If one imagines increasing $\mh$ in an otherwise fixed, triaxial
potential, the zone of chaos that extends outward from this radius
will eventually encompass the entire potential; this occurs when
the second ``perturbation parameter,'' $\mh/M_{\rm gal}$,
exceeds $\sim 10^{-2}$.
Thus at intermediate radii, in the ``zone of chaos,'' and perhaps
throughout an elliptical galaxy containing a central black hole, 
triaxiality should be difficult to maintain.

\section{MIXING AND COLLISIONLESS RELAXATION}

Stochastic motion introduces a new time scale into galactic dynamics,
the mixing time.
Mixing is the process by which a non-uniform distribution of
particles in phase space relaxes to a uniform distribution, at
least in a coarse-grained sense.
A weak sort of mixing, phase mixing\index{phase mixing}, 
occurs even in integrable potentials, 
as particles on adjacent tori gradually
move apart (Lynden-Bell 1967; Figure 8a).
Phase mixing is responsible for the fact that the coarse-grained
phase space density in relaxed integrable systems is nearly constant 
around tori.
A stronger sort of mixing takes place in chaotic systems.
Chaotic motion is essentially random in the sense that
the likelihood of finding a particle anywhere in the
stochastic region tends toward a constant value after a sufficiently
long time.
An initially compact group of stars should therefore spread out
until it covers the accessible phase space region uniformly
in a coarse-grained sense (Kandrup \& Mahon 1994; Figure 8b).
This ``chaotic mixing''\index{chaotic mixing} is 
irreversible\index{irreversibility} in the sense that an
infinitely fine tuning of velocities would be required in order
to undo its effects.
It also occurs on a characteristic time scale, the Liapunov time
associated with exponential divergence of nearby trajectories.
Phase mixing, by contrast, has no associated time scale; its rate
depends on the range of frequencies associated with orbits in the
region of interest, 
and this rate tends to zero in the case of a set of trajectories 
drawn from a single invariant torus -- a set of points on the
torus translates, unchanged, around the torus.

\begin{figure}
[ht]
\centering
\includegraphics[width=10cm, trim=0 0 0 0]{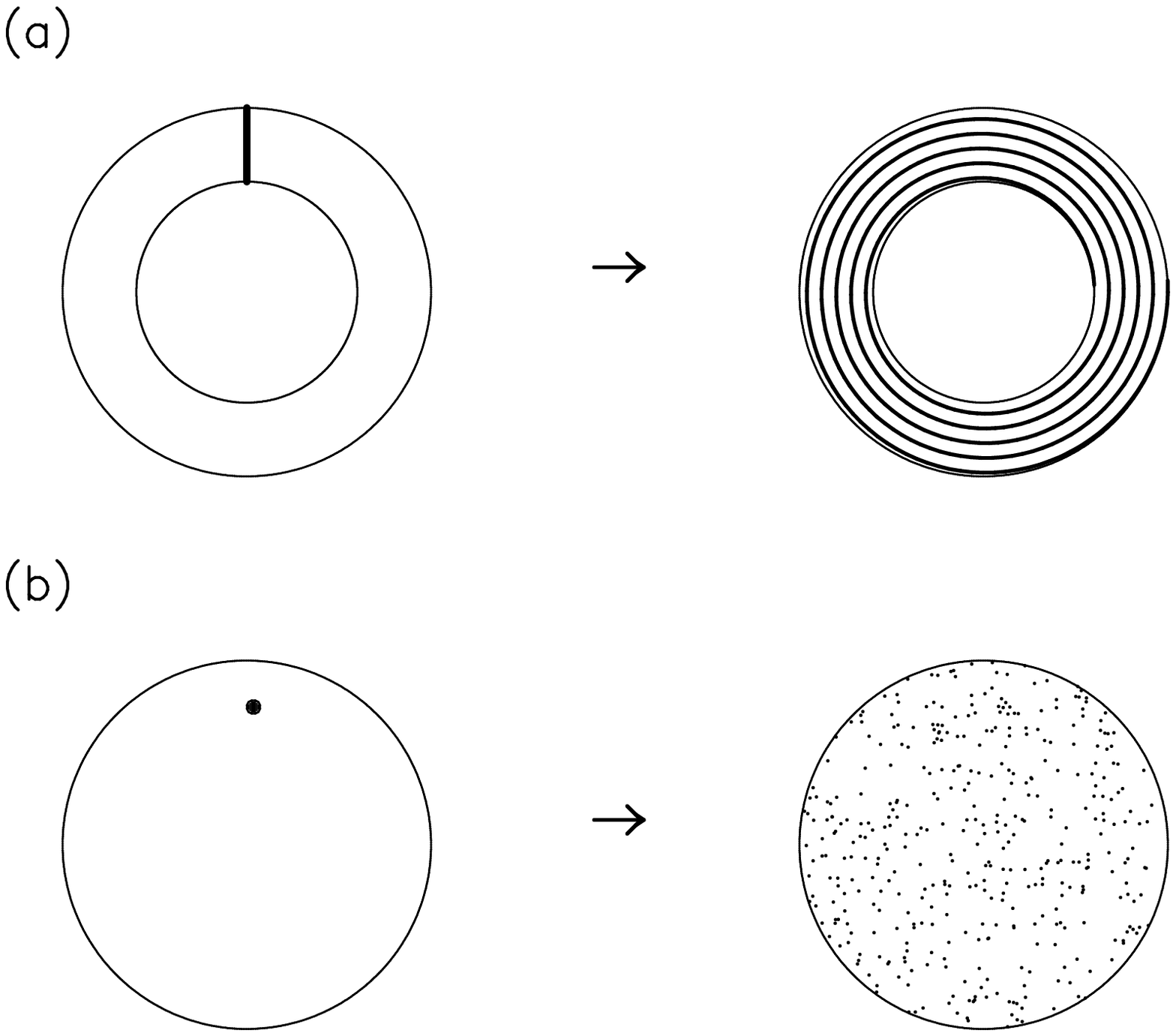}
\caption{
(a) Phase mixing vs. (b) chaotic mixing.
}
\end{figure}

Figure 9 shows examples of chaotic mixing in a triaxial potential
with a central point mass. 
Ensembles of orbits were started at rest on
an equipotential surface and integrated in tandem for several crossing
times. 
The central point had a mass $M_{\bullet} = 0.03$ in units of the
galaxy mass.
The first ensemble (a) was
begun on an equipotential surface enclosing a mass $\sim 3$ times that
of the central point; for ensembles (b) and (c) these ratios were
$\sim 7$ and $\sim 17$ respectively -- all within the ``zone of chaos.''
Mixing occurs very rapidly in these ensembles.  
At the lowest energy (ensemble a), 
the linear extent of the points in configuration space roughly
doubles every crossing time until $T\approx 4$, when the
volume defined by the equipotential surface appears to be nearly
filled.  At the highest energy (ensemble c), mixing is slower but
substantial changes still take place in a few crossing times.  The
final distribution of points at this energy  still shows some structure,
reminiscent of a box orbit.

\begin{figure}
[ht]
\centering
\includegraphics[width=13cm, trim=20 0 0 0]{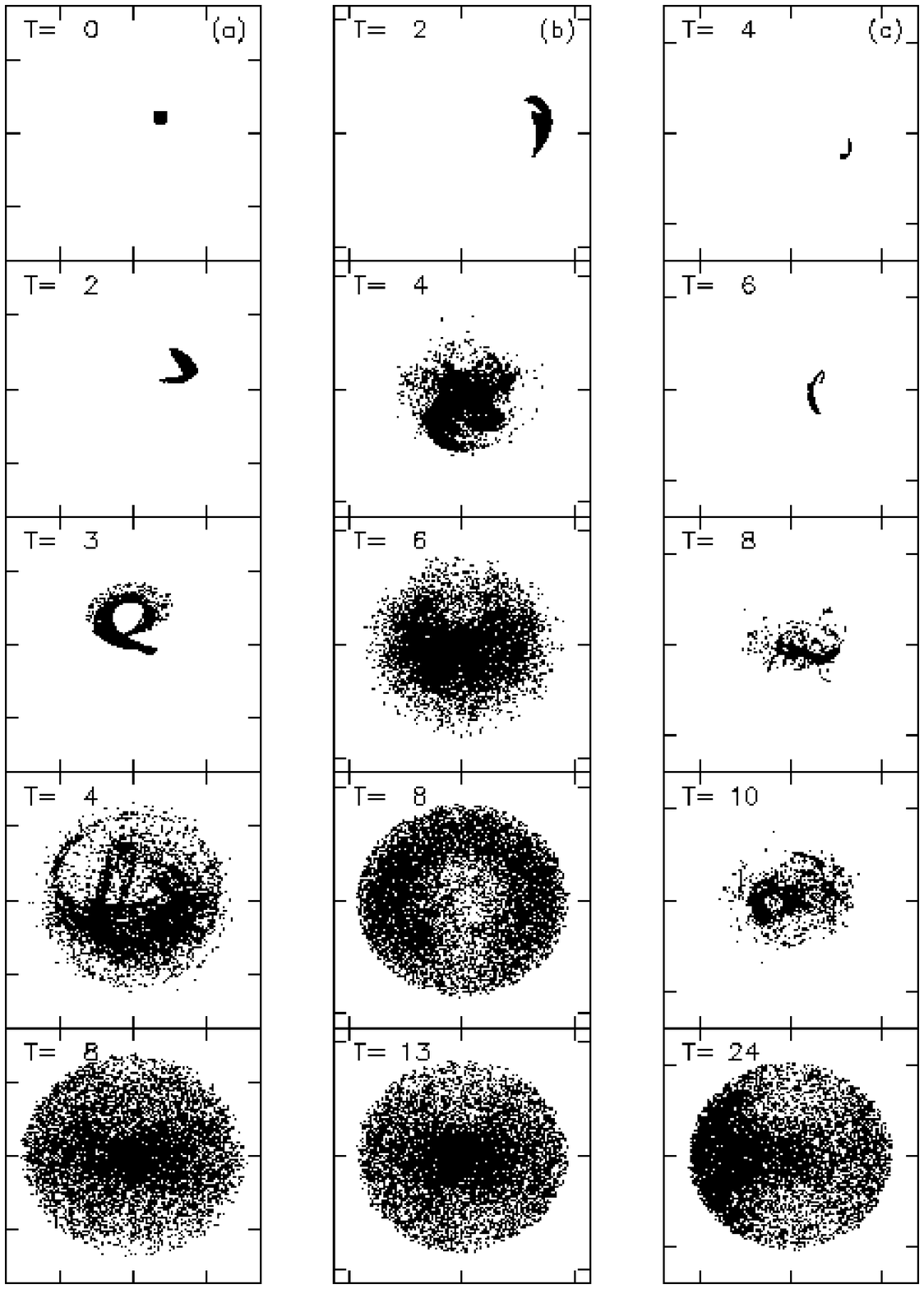}
\caption{
Mixing in a triaxial potential with a central point containing
$3\%$ of the total mass (Valluri \& Merritt 2000).
Time is in units of the local crossing time.
Ensembles of $10^4$ phase points were distributed initially
($T=0$) in patches on a equipotential surface with zero
velocity.
}
\end{figure}
  
The irreversibility of mixing flows like the ones illustrated in 
Figure 9 implies a reduction in the effective number of orbits: 
all the stochastic trajectories at a given energy are gradually
replaced by a single invariant ensemble, whose shape is typically 
not well matched to that of the galaxy (Merritt \& Fridman 1996).
If time scales for chaotic mixing are comparable to galaxy 
lifetimes, this reduction might be expected to encourage a galaxy to 
evolve away from a triaxial shape toward a more axisymmetric one, 
in which most of the orbits are tubes that avoid the destabilizing center.
Such evolution has in fact been observed in $N$-body
simulations of the response of a triaxial galaxy to the growth of a
central black hole. Merritt \& Quinlan (1998) found that a
triaxial galaxy evolves to axisymmetry in little more than the local
crossing time at each radius when the black hole mass exceeds $\sim
2.5\%$ of the total galaxy mass. 
This is about an order of magnitude larger than the typical black hole
mass ratio in real galaxies (Merritt \& Ferrarese 2001), but
Merritt \& Quinlan observed more gradual evolution even when the
mass ratio was 10 times smaller, at a rate that would imply substantial
shape changes over a galaxy lifetime.
These simulations suggest an explanation for the generally low level
of triaxiality observed in real galaxies (Bak \& Statler 2000).

\section{CHAOS IN COLLISIONAL SYSTEMS}

The discussion presented so far has assumed that galaxy potentials are 
smooth, or ``collisionless.''
In fact, the gravitational force on a star in a galaxy can be
broken up into two components: a rapidly varying component
that arises from the discrete distribution of stars,
and a smoothly varying component that arises from the large-scale
matter distribution.
The effects of the discrete component of the force relative to
the smooth component are usually assumed to scale as
$\sim \ln N/N$, the ratio of dynamical to two-body relaxation times.
For galaxies, which have $N\sim 10^{11}$, collisional effects should
therefore be unimportant.

If this were the case, it should be possible to show that the $N$-body 
trajectories go over, in the limit of large $N$, to the orbits 
in the corresponding smoothed-out potential -- i.e., that the 
equations of motion of the $N$-body problem tend to the characteristics 
of the collisionless Boltzmann equation as $N\rightarrow\infty$.
However this has never been demonstrated, 
and in fact there is an important sense
in which the equations of motion in an $N$-body system do {\it not} tend toward
the trajectories of the corresponding smooth potential in the limit of
large $N$.

\begin{figure}
[ht]
\centering
\includegraphics[width=14cm, trim=20 0 0 0]{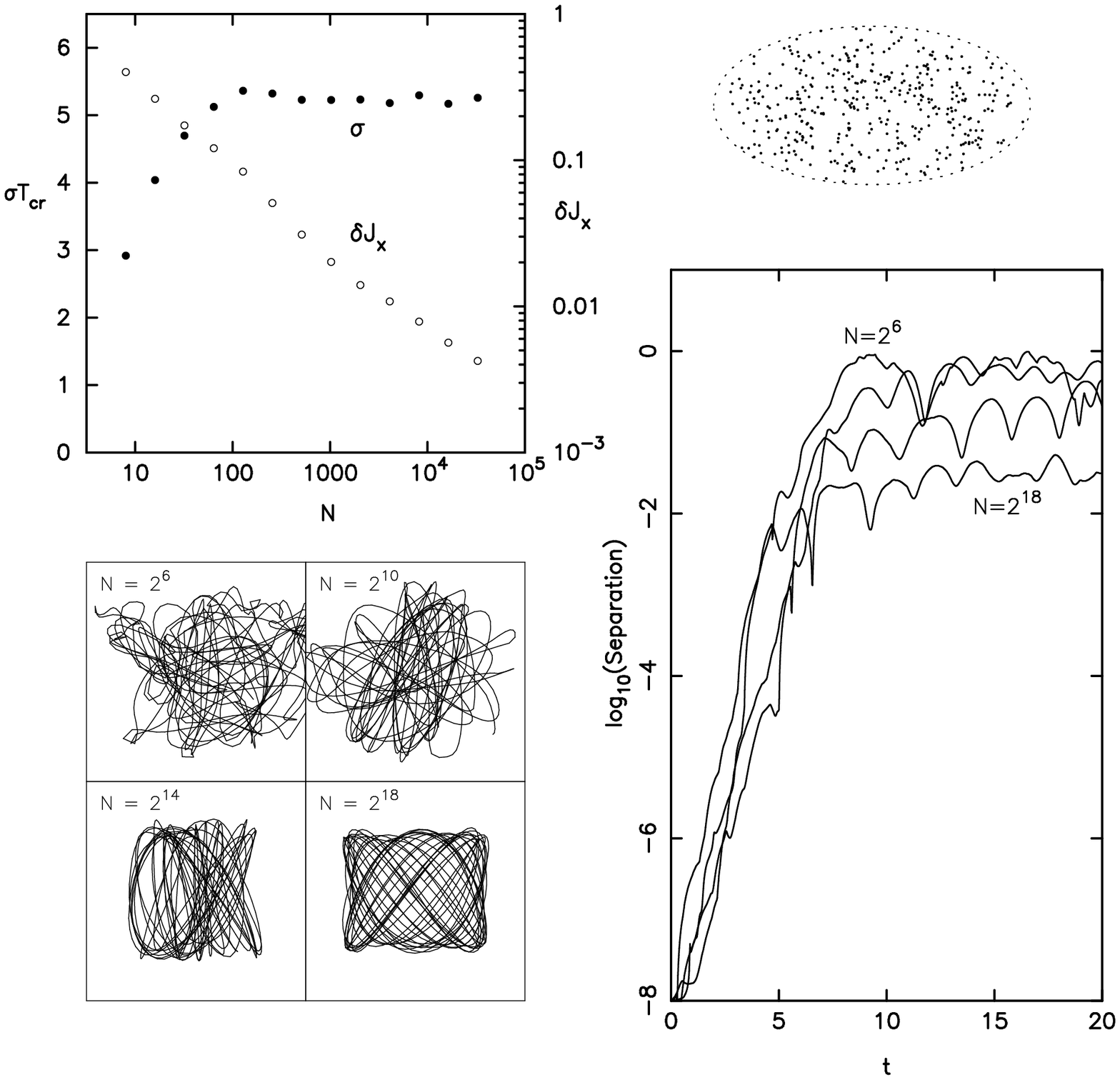}
\caption{
Evolution of orbits in a potential consisting of $N$ fixed
point masses, $m=1/N$, distributed randomly and uniformly in an
ellipsoidal volume (Valluri \& Merritt 2000) (see text).  
}
\end{figure}

This surprising statement is justified in Figure 10, which shows the
results of test-particle integrations in a potential consisting of $N$
fixed point masses distributed randomly and uniformly within a triaxial
ellipsoid.
The mass of each of the $N$ points is $m=1/N$, so that the total
mass and mean density of the ellipsoid remain constant as $N$ is varied.
In the limit $N\rightarrow\infty$, one might expect the equations of 
motion to approach those of a 3d harmonic oscillator, since the
potential of a uniform ellipsoid is quadratic in the coordinates,
$\Phi = \Phi_0\sum_i A_i x_i^2$ (Chandrasekhar 1969).
However the upper left-hand panel shows that the Liapunov exponents
$\sigma$ of orbits in the $N$-body potential do {\it not} tend to zero with 
increasing $N$.
Instead, the instability\index{instability} time scale appears to r
each a roughly 
constant value (expressed as a fraction of the crossing time $T_{cr}$)
for $N\gap 10^3$.
Furthermore the instability time scale is very short, 
a fraction of the crossing time!

The generic instability of the $N$-body problem was first noted
by Miller (1964), who calculated the time evolution of the separation
between two $N$-body systems with slightly different initial conditions.
He defined this separation as
\beq
\Delta(t) = \left[\sum\left({\bf x}_2-{\bf x_1}\right)^2 +
		\sum\left({\bf v}_2-{\bf v_1}\right)^2\right]^{1/2}
\eeq
with ${\bf x}_1$ and ${\bf x}_2$ the $N$ configuration-space 
coordinates in $N$-body systems $1$ and $2$ and 
${\bf v}_1$ and ${\bf v}_2$ the velocities; 
the summations extend over all the particles.
Miller found, for $4\leq N\leq 32$, that $\Delta$ grew roughly 
exponentially with a characteristic time scale that was a fraction
of the crossing time, as in the fixed $N$-body problem of Figure 10.

What are the physical implications of this generic instability?
Several suggestions have been made.
Gurzadyan \& Saviddy (1986), who first investigated the large-$N$ 
dependence of the instability using an idealized model, 
suggested that the exponential divergence implies chaotic mixing on 
a similar time scale,
and hence that stellar systems should relax much more rapidly than
implied by the standard Chandrasekhar formula.
Heggie (1991) disagreed, but suggested that the use of smooth 
potentials for approximating galaxies would need to be abandoned,
at least for studies of orbital instability.
Kandrup (1998) suggested that -- while individual orbits may always
be exponentially unstable -- ensembles of $N$-body systems 
might behave, on average, as if the potential were smooth.

Figure 10 suggests an even stronger way in which the motion
goes over to that of the collisionless problem as $N\rightarrow\infty$.
The open circles in the upper left-hand panel of that figure
show a second measure of the orbital evolution: 
the rms variation, over 20 orbital periods, 
of the action $J_x$ for each ensemble of orbits.
Contrary to the behavior of the Liapunov exponents, 
the average changes
in the actions tend uniformly to zero as 
$N$ is increased -- in other words, the orbits approach more and more
closely, in their {\it macroscopic} behavior, to that of 
integrable orbits even though they remain {\it locally} unstable 
(as measured by the Liapunov exponents) to a
degree that is nearly independent of $N$.
Plots of the trajectories of some typical orbits (lower left panel)
confirm this interpretation.
These results suggest the way in which trajectories in the $N$-body 
problem tend toward those in the corresponding smooth potential:
as $N$ is increased, orbits are confined more and more strongly to
narrow regions of phase space around the invariant tori of the smooth
potential.

It is remarkable that orbits can be extremely unstable locally, 
as measured by their Liapunov exponents, 
and yet behave macroscopically in a way that is
essentially identical to that of regular orbits.
Apparently, the exponential growth of perturbations must saturate at some
finite amplitude, and this saturation amplitude must be a decreasing 
function of $N$.
The lower right-hand panel of Figure 10 verifies this conjecture
for a few pairs of orbits with nearly identical initial conditions.
The early divergence takes place at a rate that is independent of
$N$, but for large $N$, the separation saturates at a value that is
much smaller than the size of the system.
These pairs of orbits act as if they are confined to the same, 
restricted region of phase space; 
saturation occurs when the separation between them is of
order the width of this region.  
The fact that the exponential
divergence saturates sooner for larger $N$ suggests that the width of
the confining regions decreases with increasing $N$.

These results suggest that collisional relaxation
in stellar systems is intimately connected with the evolution
of orbits under conditions of weak chaos, i.e., with Arnold
diffusion.
This connection would be a fruitful topic for future study.

Some of the work presented here was first published in collaboration
with M. Valluri.
I am grateful for her permission to reproduce the work here.
The preparation of this review was supported by NSF grant 
AST-0071099 and by NASA grants NAG 5-2803 and NAG 5-9046.

\section*{References}

\frenchspacing
\begin{small}

\reference{Arnold, V. I. 1963, {\em Russian Mathematical Surveys}, 
	{\bf 18}, 85}

\reference{Bak, J. \& Statler, T. 2000, {\em AJ}, {\bf 120}, 110}

\reference{Barnes, J. 1996, in {\em The Formation of
	Galaxies}, Proceedings of the V Canary Islands Winter School
	of Astrophysics, ed. C. Mu\~noz-Tu\~n\'on
	(Cambridge: Cambridge University Press), 399}

\reference{Bicknell, G. V., Koekemoer, A.,
	Dopita, M. A. \& O'Dea, C. P. 1997, in {\em The Second Stromlo
	Symposium: The Nature of Elliptical Galaxies}, A.S.P. Conf. 
	Ser. Vol. 116, eds. M. Arnaboldi, G. S. Da Costa \& P. Saha) 
	(Provo: ASP), 432}

\reference{Binney, J. \& Kumar, S. 1993, {\em MNRAS}, {\bf 261}, 584}

\reference{Binney, J. \& Spergel, D. 1982, {\em ApJ}, {\bf 252}, 308}
 
\reference{Binney, J. \& Spergel, D. 1984, {\em MNRAS}, {\bf 206}, 159}
 
\reference{Boozer, A. H. 1982, {\em Phys. Fluids}, {\bf 25}, 520}
 
\reference{Carpintero, D. D. \& Aguilar, L. A. 1998, {\em MNRAS}, 
	{\bf 298}, 1}

\reference{Chandrasekhar, S. 1969, {\em Ellipsoidal Figures of
	Equilibrium} (New York: Dover)}

\reference{Chapman, S., Garrett, B. C. \& Miller, W. H. 1976, 
	{\em J. Chem. Phys.}, {\bf 64}, 502}

\reference{Crane, P. et al. 1993, {\em AJ}, {\bf 106}, 1371}

\reference{de Zeeuw, P. T. \& Lynden-Bell, D. 1985, {\em MNRAS}, 
{\bf 215}, 713}

\reference{Eaker, C. W., Schatz, G. C., De Leon, N. \& Heller, E. J. 1984, 
	{\em J. Chem.\ Phys.}, {\bf 81}, 5913}
 
\reference{Ferrarese, L. et al. 1994, {\em AJ}, {\bf 108}, 1598}

\reference{Ferrarese, L. \& Merritt, D. 2000, {\em ApJ}, {\bf 539}, L9}

\reference{Franx, M., Illingworth, G. D. \& de Zeeuw, P. T. 1991, 
	{\em ApJ}, {\bf 383}, 112}

\reference{Franx, M., Illingworth, G. D. \& Heckman, T. M. 1989, {\em ApJ}, 
	{\bf 344}, 613}

\reference{Gerhard, O. E. \& Binney, J. J. 1985, {\em MNRAS}, {\bf 216} 467}

\reference{Goldstein, H. 1980, {\em Classical Mechanics\/} 2nd ed.\ (Reading:
Addison-Wesley)}

\reference{Goodman, J., Heggie, D. C. \& Hut, P. 1993, {\em ApJ}, {\bf 415}, 
	715}.

\reference{Goodman, J. \& Schwarzschild, M. 1981, {\em ApJ}, {\bf 245}, 1087}

\reference{Guerra, D. V. \& Ratcliff, S. J. 1990, {\em ApJ}, {\bf 348}, 127}

\reference{Gurzadyan, V. G. \& Savvidy, G. K. 1986, {\em A\&A}, {\bf 160}, 
	203}.

\reference{Heggie, D., 1991, in {\em Predictability, Stability, and 
	Chaos in $N$-Body Dynamical Systems}, ed. A. E. Roy
	(Plenum Press, New York) p. 47.}

\reference{Kaasalainen, M. 1994, {\em MNRAS}, {\bf 268}, 1041}

\reference{Kaasalainen, M., \& Binney, J.1994a, {\em MNRAS}, {\bf 268}, 1033}

\reference{Kaasalainen, M., \& Binney, J. 1994b, {\em Phys. Rev. Lett.}, {\bf 73}, 2377}

\reference{Kandrup, H. E. 1998, in {\em Long-Range Correlations in 
	Astrophysical Systems}, ed. J. R. Buchler, J. W. Dufty
	\& H. E. Kandrup, {\em Ann. N Y Acad. Sci.}, {\bf 848},
	28}
	
\reference{Kandrup, H. E. \& Mahon, M. E. 1994, {\em Phys. Rev. E}, 
	{\bf 49}, 3735} 

\reference{Kormendy, J. 1985, {\em ApJ}, {\bf 292}, L9}

\reference{Kormendy, J. \& Richstone, D. O. 1995, {\bf ARA\&A}, {\em 33}, 581} 

\reference{Kuo-Petravic, G., Boozer, A. H., Rome, J. A. \& Fowler, R. H.
1983, {\em J. Comp. Phys.}, {\bf 51}, 261}

\reference{Kuzmin, G. G. 1973, in {\em The
	Dynamics of Galaxies and Star Clusters}, ed.
	G. B. Omarov (Nauka, Alma Ata)}

\reference{Landau, L. D. \& Lifshitz, E. M. 1976, {\em Mechanics\/} 3rd
	ed.\	(Oxford: Pergamon)}
 
\reference{Laskar, J. 1988, {\em AAp}, {\bf 198}, 341}
 
\reference{Laskar, J. 1990, {\em Icarus}, {\bf 88}, 266}
 
\reference{Laskar, J. 1993, {\em Physica D}, {\bf 67}, 257}
 
\reference{Laskar, J. 1996, in {\em Hamiltonian Systems with Three or More
Degrees of Freedom\/} NATO-ASI, eds.\ C. Simo \& A. Delshams (Dordrecht:
Kluwer), REFERENCE??}
 
\reference{Lichtenberg, A. J. \& Lieberman, M. A. 1992,
	{\em Regular and Chaotic Dynamics} (New York:
	Springer)}

\reference{Lynden-Bell, D. 1967, {\em MNRAS}, {\bf 136}, 101}

\reference{Martens, C. C. \& Ezra, G. S. 1985, {\em J. Chem. Phys.}, {\bf 83}, 2990}

\reference{McGill, C. A., \& Binney, J. 1990, {\em MNRAS}, {\bf 244}, 634}

\reference{Merritt, D. 1980, {\em ApJS}, {\bf 43}, 435}

\reference{Merritt, D. \& Ferrarese, L. 2001, {\em MNRAS}, {\bf 320}, L30}

\reference{Merritt, D. \& Fridman, T. 1996, {\em ApJ}, {\bf 460}, 136}

\reference{Merritt, D. \& Quinlan, G. 1998, {\em ApJ}, {\bf 498}, 625}

\reference{Merritt, D. \& Valluri, M. 1999, {\em AJ}, {\bf 118}, 1177}

\reference{Miller, R. H. 1964, {\em ApJ}, {\bf 140}, 250}.

\reference{Miralda-Escud\'e, J. \& Schwarzschild, M. 1989, {\em ApJ}, 
	{\bf 339}, 752}

\reference{Papaphilippou, Y. \& Laskar, J. 1996, {\em A\&A}, {\bf 307}, 427}

\reference{Papaphilippou, Y. \& Laskar, J. 1998, {\em A\&A}, {\bf 329}, 451}

\reference{Percival, I. C. 1974, {\em J. Phys. A}, {\bf 7}, 794}

\reference{Percival, I. C. 1977, {\em J. Phys. A}, {\bf 12}, 57}

\reference{Poincar\'e, H. 1892, {\em Les M\'ethodes Nouvelle de la M\'ecanique
	C\'eleste Tome I.} (Paris: Gauthier-Villars), ch. 3}

\reference{Poon, M. \& Merritt, D. 2001, {\em ApJ}, in press 
	(astro-ph/0006447)}

\reference{Ratcliff, S. J., Chang, K. M., \& Schwarzschild, M. 1984, 
	{\em ApJ}, {\bf 279}, 610}

\reference{Reiman, A. H. \& Pomphrey, N. 1991, {\em J. Comp. Phys.}, {\bf 94}, 
	225}
 
\reference{Ryden, B. S. 1996, {\em ApJ}, {\bf 461}, 146}

\reference{Sambhus, N. \& Sridhar, S. 2000, {\em Apj}, {\bf 542}, 143}

\reference{Sanders, D. B. \& Mirabel, I. F. 1996, {\em ARAA}, {\bf 34}, 749} 

\reference{Schechter, P. L. \& Gunn, J. E. 1979, {\em ApJ}, {\bf 229}, 472}

\reference{Schwarzschild, M. 1979, {\em ApJ}, {\bf 232}, 236}

\reference{Schwarzschild, M. 1982, {\em ApJ}, {\bf 263}, 599}

\reference{Shlosman, I., Begelman, M. C. \& Frank, J. 1990, {\em Nature}, 
	{\bf 345}, 679}

\reference{Sridhar, S. \& Touma, J. 1997, {\em MNRAS}, {\bf 287}, L1}

\reference{Tremblay, B. \& Merritt, D. 1995, {\em AJ}, {\bf 110}, 1039}

\reference{Tremblay, B. \& Merritt, D. 1996, {\em AJ}, {\bf 111}, 2243}

\reference{Valluri, M. \& Merritt, D. 1998, {\em ApJ}, 506, 686}

\reference{Valluri, M. \& Merritt, D. 2000, in {\em The Chaotic Universe},
	ed. V. G. Gurzadyan \& R. Ruffini (Singapore: World Scientific),
	229}

\reference{van der Marel, R. P., Cretton, N., de Zeeuw, P. T. \& Rix, H. W. 
	1998, {\em ApJ}, {\bf 493}, 613}

\reference{Wachlin, F. C. \& Ferraz-Mello, S. 1998, {\em MNRAS}, {\bf 298}, 22}
 
\reference{Warnock, R. L. 1991, {\em Phys. Rev. D}, {\bf 66}, 1803}
 
\reference{Warnock, R. L. \& Ruth, R. D. 1991, {\em Phys. Rev. Lett.}, {\bf 66}, 990}
 
\reference{Warnock, R. L. \& Ruth, R. D. 1992, {\em Physica D}, {\bf 56}, 188}

\end{small}
\end{document}